\documentstyle[12pt]{article}
\begin{document}
\baselineskip=0.89cm
\vspace{4cm}
\begin{center}
{\bf Composite-Fermion Picture for the Spin-Wave Excitation\par
in the fractional quantum Hall system}\par
\par
\ \\
T. Nakajima and H. Aoki\par
{\it Department of Physics, University of Tokyo, Hongo, Tokyo 113, Japan}\par
\end{center}
\vspace{0.4cm}
\par
\noindent {\bf Abstract :}\hspace{4pt}  Spin-wave excitation mode from the
spin-polarized ground state in the fractional quantum Hall liquid with odd
fractions ($\nu=1/3,1/5$) numerically obtained by the exact diagonalization of
finite systems is
shown to be accurately described,
for wavelengths exceeding the magnetic length,
in terms of
the composite-fermion mean-field approximation for the spin-wave (magnon)
theory formulated in the spherical geometry.
This indicates that the composite picture extends to excited states, and also
provides the spin stiffness in terms of peculiar exchange interactions.

\ \\


\newpage

Although it is more than a decade since the fractional quantum Hall
(FQH) effect\cite{qhe} was discovered and identified as a novel prototype of
strongly correlated electron system, a new way of looking at the
problem is now being explored in terms of the composite-fermion
picture.
 There is a mounting evidence that the composite picture, which was first
proposed by Jain\cite{jain} as an alternative to the Haldane-Halperin hierarchy
\cite{haldane83,halperin84},
is surprisingly
adequate.
 The composite picture asserts that a quantum Hall liquid of electrons in the
external magnetic field $B$, which corresponds to $\nu ^{-1} = 2m + 1$ flux
quanta per electron ($m$: an integer), is equivalent, in a mean-field sense, to
 a liquid of composite fermions each carrying $2m$ flux quanta in the effective
magnetic field $B_{\rm eff} = B/(2m + 1)$ that corresponds to $\nu^{-1} = 1$.

 Following the Chern-Simons (CS) gauge-field theory\cite{cs}, Halperin, Lee and
Read have indeed shown that the spinless system should behave like a system of
composite-fermions for which $\nu = 1/2$ corresponds to zero magnetic
field.\cite{hlr}
The fermionic CS approach has been extended to discuss the dynamical response
and effective mass.\cite{simon}
 Recent experiments by Willett {\em et al} \,and by Kang {\em et al} \,confirm
the Fermi-liquid-like behaviors around $\nu = 1/ 2$.\cite{willett,kang}
 Du {\em et al} \,showed that the transport property in the FQHE state around
$\nu = 1/2$ resembles the usual Shubnikov-de Haas oscillation with a
well-defined energy gap $\Delta$,\cite{du} for which
Leadley {\em et al} \,have subsequently estimated the effective mass from the
temperature-dependence of the Shubnikov-de Haas oscillation.\cite{leadley}
Goldman et al report on magnetic focusing.\cite{goldman}
These results are rather surprising, since there is no apriori reason why the
composite-particle picture
 should be a good approximation.

 Now, we consider that a true test for a many-body theory is its ability to
describe excited states.
 For odd-fraction Landau-level fillings, $\nu = 1/ (2m+1)$, the ground state
(Laughlin's quantum liquid) is fully spin-polarized even when the Zeeman energy
is neglected.
 Then the low-lying excitations are the collective spin-wave mode that restores
the broken SU(2) symmetry (rotational symmetry in spin
space).\cite{kallin,rasolt,sondhi}
Hence the problem may also be regarded as looking into a two-component (spin
up/down) FQH system,
while the usual practice in studying the odd fractions is to just ignore the
spin.
The spin degrees of freedom are especially fascinating, since a most drastic
effect of electron correlation in the ordinary correlated system (e.g., the
Hubbard model) is the spin state, which is thought of as a manifestation of the
exchange interaction in the appropriate basis.
This usually involves the competition of the kinetic and interaction energies,
which causes intriguing anomalies in the spin wave dispersion in the Hubbard
model that becomes spin-polarized due to the electron
correlation.\cite{kusakabe}
By contrast, the Landau-level filling alone dominates the intra-Landau level
excitations such as the spin wave in the FQH system
due to the quenched kinetic energy.
Thus a prominent question is: can we extend the the composite
picture to ferromagnetic spin-wave excitation spectra?

The spin wave has recently been experimentally observed by Pinczuk {\em et al}
\, in which a sharp peak at the energy of the Zeeman splitting, $g \mu_{B} B$,
detected with the inelastic light scattering at $\nu = 1/3$ is assigned to the
long wavelength ($k\simeq 0$) spin-wave excitation.\cite{pinczuk}

 Here we propose an application of the composite-particle picture to the
spin-wave (magnon) theory for the FQH system, which is formulated in the
spherical geometry to exploit the rotational symmetry\cite{issp}.
 The result is compared with the exact spin-wave excitation numerically
obtained from the diagonalization of finite FQH systems.
We shall show that the spin-wave excitation spectrum for $\nu = 1/( 2m + 1 )$
can be explained surprisingly accurately by a composite-fermion picture with a
mean-field approximation unless the wavelength is smaller than the magnetic
length.
One quantitative outcome is that this approach enables us to determine the spin
stiffness in the FQH system, in which the magnitude of the spin-spin coupling
is difficult to conceive in a conventional way.

 We start from the spin-wave excitation spectrum for the flat geometry at $\nu
= 1$, which has been exactly given by Kallin and Halperin \cite{kallin} as
\begin{equation}
\omega ( k ) - g \mu _B B  =
 \frac{1}{2 \pi} \int_{0}^{\infty} dq \ q V(q)
 \Bigl[ 1 - J_0 ( k q \ell^2 ) \Bigl]
 e^{-(q \ell)^2/ 2}, \label{eqn:kh}
\end{equation}
 where $V(q) = 2 \pi e^2 / \epsilon q$ the Fourier transform of the Coulomb
interaction with $\epsilon$ being the dielectric constant, $J_0 ( z )$ Bessel's
function, and $\ell = \sqrt{c \hbar/e B}$ the magnetic length.

If we now turn to the spherical geometry, everything can be written in terms of
angular momentum quantum numbers.
 As we stereographically map the flat system to the spherical one, the
translational symmetry is translated into the rotational symmetry, so that
the wavenumber $k$ and total angular momentum $L$ are related by $k = L/R$,
where $R$ is the radius of the sphere.
  When the total magnetic flux going out of the sphere is $2S$ (an integer due
to Dirac's condition) times the flux quantum, the radius of the sphere becomes
$R = \ell \sqrt{S}$, while the relation to $\nu$ is $2S = \nu^{-1}N -$ integer
with $N$ being the number of electrons.
  There the creation operator for the magnon with angular momentum $L$ and its
$z$ component $M$ is given by
 $C^{\dagger}_{L M} = \sum_{j, k}\ (-1)^{S-k} \langle S,j;S,-k|LM\rangle
a^{\dagger}_{j \downarrow} a_{k \uparrow},$
where $\langle S,j;S,-k|LM\rangle$ is the Clebsch-Gordan coefficient,
$a^{\dagger}_{j \sigma}$ is the creation operator for $j$-th spatial orbit with
spin $\sigma$.

The spin-wave excitation spectrum, $\omega _L$, which is a function of $L$ in
the spherical geometry, for $\nu = 1$ requires a tedious calculation, but the
result is rather elegant in that it is given in terms of Wigner's $6j$-symbol,
familiar in the nuclear physics, as
\begin{equation}
 \omega _L - g \mu _B B = \sum _{J = 0}^{2S}\ ( 2J+1 )\ (-1)^{2S-J}\ V_J\
\biggl[\ \frac {1}{2S+1}\ -\ ( -1 )^{2S-J}\
\left\{
\begin{array}{ccc}
S & S & L\\
S & S & J
\end{array}
\right\}\ \biggl] ,\label{eqn:sw1}
\end{equation}
 where $L (= 0, 1, \cdots, 2S$) is the total angular momentum,
 $\{^{S S L}_{S S J}\}$ is the $6j$-symbol, and $V_J$ is the Haldane
pseudopotential for the relative angular momentum $2S-J$
\cite{haldane83,qhehal}.
 Since $\{ ^{S S \,\mbox{{\footnotesize 0}}} _{S S J} \} = (-1)^{2S-J}/
(2S+1)$, the excitation energy at $k = L/ R = 0$ satisfies the relation $\omega
_0 = g \mu _B B$, for any inter-particle interaction (\{$V_J$\}), which
guarantees
Larmor's theorem.

Now we apply Jain's composite-fermion picture by attaching $2m$ flux quanta to
each electron in the $\nu=1$ state.
 When we attach $2m$ flux quanta to each electron, the relative angular
momentum, $n$, between two electrons translates into the relative angular
momentum $n - 2m$ between composite fermions, since an extra phase factor
$e^{2m \theta i}$ appears in the wavefunction of the relative
motion\cite{halperin82} (as often described in terms of the CS theory in the
literature).
 Since the field is reduced to $B_{\rm eff} = B/ ( 2m + 1 )$, the magnetic
length $\ell$ changes into $\tilde{\ell} = \sqrt{2m + 1}\,\ell$ in a mean-field
sense.

We are now in position to formulate the spin-wave excitation at $\nu = 1/ ( 2m
+ 1 )$.
The advantage of working in the spherical geometry is that
the transformation into the composite-particle picture is simply given by
\begin{equation}
2\tilde{S} = 2S/(2m + 1) = N - 1,\ \ \
\tilde{V}_{2\tilde{S}-(n-2m)}/\ \frac{\ e^2}{\epsilon \tilde{\ell}} \,= \,
V_{2S-n}/\ \frac{\ e^2}{\epsilon \ell}  \label{eqn:cf1},
\end{equation}
where $V_{2S-n}$ is the pseudopotential with the relative angular momentum $n =
2S - J$.
 If we plug this transformation into the `6$j$-formula', eqn(\ref{eqn:sw1}), we
finally arrive at the desired expression for the spin-wave excitation spectrum
for $\nu = 1/ (2m+1)$ in the composite picture as
\begin{equation}
 \omega _L - g \mu _B B = \sum _{J = 0}^{2\tilde{S}}\,( 2J+1
)\,(-1)^{2\tilde{S}-J}\ \tilde{V}_J\ \biggl[\ \frac {1}{2\tilde{S}+1}\ -\ ( -1
)^{2\tilde{S}-J}\,
\left\{
\begin{array}{ccc}
\tilde{S} & \tilde{S} & L\\
\tilde{S} & \tilde{S} & J
\end{array}
\right\}\ \biggl], \label{eqn:cfmfa}
\end{equation}
 where the range of $L$ now reduces to $L = 0, 1, \cdots, 2\tilde{S}$.

 We now turn to the numerical results
for the low-lying excitations in the spherical geometry for a $6$-electron
system at $\nu = 1/ 3$ and for a $5$-electron system at $\nu = 1/ 5$ in Fig.1,
 where we show both the one-spin-flip excitations (with the change in the total
spin $\Delta S_{\rm tot} = -1$\,) and the charge (\,$\Delta S_{\rm tot} = 0$\,)
excitations.
 In the spin-wave mode, which is the lowest branch in the excitation,
we immediately recognize that the states in these finite systems appear only in
the range $0 \leq L \leq 2\tilde{S} = N - 1$, while naively there is no reason
why the states should not extend for $0 \leq L \leq 2S= ( 2m + 1 ) ( N - 1 )$.
In fact, higher-energy spin excitations do indeed exist for larger $L$ in
Fig.1.
 We attribute this truncation from $( 2m + 1 ) ( N - 1 )$ to $( N - 1 )$ to the
fact that the original system at $\nu = 1/ ( 2m + 1 )$ may be mimicked by a
system of composite fermions with $\nu = 1$ for the spin-wave excitation.

If we look at the spin-wave dispersion curve in Fig.1, the prediction from the
composite-fermion mean-field theory (CFMFA),
eqn(\ref{eqn:cfmfa}), exhibits an excellent
agreement with the exact result up to the wavenumber $k \sim \ell ^{-1}$.
 The exact result starts to deviate from the CFMFA for larger $k$, which
implies that the effect of fluctuations from the mean CS field becomes
appreciable for large-wavenumber excitations.
 The fluctuations should also dominate the higher-energy spin excitations above
the spin-wave mode, which have no counterparts for $\nu = 1$.

 The result that both the number of discrete modes in a finite system and their
dispersion agree with the prediction from the CFMFA confirms the picture that
the composite-fermion picture is valid not only for the ground state but also
for the spin-wave excitation in the long-wavelength region.
 This is the key message of the present Letter.

 To look into the size dependence of the results for the spin-wave excitation,
 we have calculated the CFMFA result for a larger $51$-electron spherical
system at $\nu = 1/ 3$ in Fig.2.
For odd fractions, there exists a theory for the spin-wave excitation spectrum
by Rasolt {\em et al}\cite{rasolt} in the single-mode approximation (SMA). This
gives
$\omega ( k )$ as
\begin{equation}
\omega (k) - g \mu _B B = \frac{1}{2 \pi} \int_{0}^{\infty} dq \ q V(q)
 \Bigl[ 1 - J_0 ( k q \ell^2 ) \Bigl]
 \Bigl[ 1 - S( q ) \Bigl], \label{eqn:smasw}
\end{equation}
with $S( q )$ being the static structure factor for the fully-polarized ground
state, which agrees with
numerical results obtained for finite rectangular\cite{yoshioka} or
spherical\cite{rezayi} systems
for small wavenumbers within the finite-size correction.
Thus we have also plotted the SMA result for an infinite (flat)
system\cite{SMAMC} along with the exact diagonalization result for a
$5$\,($6$)-electron system.
  We can see that all of the exact, CFMFA and SMA results agree with each other
for wavenumbers up to $k \sim \ell^{-1}$.
 The exact result for the finite system for $k < \ell ^{-1}$ is slightly larger
than other results.  This we consider comes partly from a finite-size
correction: the Haldane pseudopotential becomes larger\cite{qhehal} for finite
systems than that for an infinite system (by about $5 \%$ for the present
size).
Thus the agreement of the CFMFA result persisits for larger systems.
 For $k > \ell^{-1}$
the SMA result, which is also intended for long-wavelengths,\cite{rasolt} the
finite-system result, and the CFMFA result start to deviate from each other.

Intuitively the reason why the composite picture is good may be traced back to
the `correlation hole': the probability of two repulsively-interacting
electrons coming to the distance closer than $\sim \ell$ is so small that, for
an electron in a long-wavelength spin wave, whether the flux is uniform or
coalesced into a filament for the other electrons is unimportant.

 As for the stiffness of the spin wave, this can be estimated by transforming
the electron-electron interaction in the Kallin-Halperin formula, eqn
(\ref{eqn:kh}), via eqn(3) to have
\begin{equation}
V(q) = 2 \pi \ell^2 \sum _{n = 0}^{\infty} 2 \,V_n L_n (q^2 \ell^2)
\longrightarrow
2 \pi \ell^2 \sum _{n = 0}^{\infty} \frac {2 \,V_{n+2m}}{\sqrt{2m+1}} \,L_n
(q^2 \ell^2),
\end{equation}
 where $L_n (z)$ is Laguerre's polynomial \cite{qhehal}.
 The stiffness, $D$, of the spin-wave defined by $\omega (k)/ (e^2/\epsilon
\ell)\,=D(k \ell)^2$ \,for small $k$ is thus expressed as an (asymptotic)
expansion involving
 peculiar `exchange integrals' in the angular-momentum space.
 Specifically, we can see the spin stiffness significantly decreases as we go
from $\nu = 1$ down to 1/3, 1/5, ..., since $V(q)$ becomes progressively
reduced.

 As for the spin excitations other than the one-spin flips, we can show that
all the low-lying excitations at $\nu = 1$ can be entirely interpreted in terms
of the multiplets of weakly-interacting magnons where the interaction is
attractive at short distances.\cite{issp}
 The attraction appears in the two-magnon states, for which the state having
the largest possible total angular momentum has the lowest energy within a
multiplet.
 This is consistent with Rezayi's observation that the two-spin-flip mode has a
lower energy than that of the one-spin-flip mode at $\nu = 1$ \cite{rezayi91}.
 Sondhi {\em et al} \,went on to discuss the many-spin flips in analogy with a
skyrmion.\cite{sondhi}
 It is an interesting problem to ask the applicability of the composite picture
to these multi-spin-flip excitations.
  We also notice in Fig.1 here that the roton-like charge-excitation
mode\cite{girvin} exists as well, identified as a $\Delta S_{\rm tot} = 0$
dispersion with an energy gap.
 Extension of the composite-particle picture to charge modes is also an
interesting future problem.

 We are grateful to Prof. D. Yoshioka and Koichi Kusakabe for valuable
discussions, and also
to Takahiro Mizusaki, Michio Honma and Tsutomu Sakai for helpful discussions on
the numerical formulae involving $3j$-symbols.
 The numerical calculations were done on HITAC S3800 in the Computer Centre,
the University of Tokyo.
 This work was in part supported by a Grant-in-Aid from the Ministry of
Education, Science and Culture, Japan.
\par

\newpage
%
\noindent {\bf Figure captions}

\

\noindent {\bf Fig.1} The excitation spectrum for a FQH system of spin $1/ 2$
electrons, which comprises one-spin-flip excitations with $\Delta S_{\rm tot} =
-1$ (open circles\,) and charge excitations with $\Delta S_{\rm tot} = 0$ (open
squares\,), is shown for (a) a $6$-electron system at $\nu = 1/ 3$ and (b) a
$5$-electron system at $\nu = 1/ 5$ (note a difference in vertical scales).
 The result for the spin-wave excitation in the composite-fermion mean-field
approximation for the same number of electrons is shown by solid triangles.
The spin-wave and roton excitations are respectively connected by a curve as a
guide to the eye.
\par
\ \\
\noindent {\bf Fig.2} The spin-wave excitation spectrum for a FQH system of
spin $1/ 2$ electrons at $\nu = 1/ 3$ is shown.
 The exact result for a $5$\,($6$) electron system is indicated by solid (open)
circles.
 The result for a $51$-electron system in the composite-fermion mean-field
approximation in the spherical geometry (solid triangles) and the single-mode
approximation for a infinite (flat) system (solid curve) are also indicated.

\begin{thebibliography}{99}
\bibitem{SMAMC} In the SMA calculation using the eqn(5),
 the static structure factor $S( q )$ for the ground state is needed.
 Here we have employed Girvin's fitting to the Monte-Carlo result
\cite{girvin}.
\end{thebibliography}
\end{document}